\documentclass[draftcls, onecolumn, 12pt]{IEEEtran}
\usepackage{cite}
\usepackage{epsfig,amssymb}
\usepackage{graphics}
\usepackage{amsmath}
\usepackage{verbatim}
\usepackage{bbm}
\usepackage{psfrag}
\usepackage[usenames]{color}

\newcommand{\nr}{n_{\mathrm{R}}}
\newcommand{\Gr}{\vec{G}_{\mathrm{r}}}    
\newcommand{\Hr}{\vec{H}_{\mathrm{r}}}    
\newcommand{\nt}{n_{\mathrm{T}}}

\newcommand{\p}[1]{\mathop{\mbox{\it p} } }

\renewcommand{\vec}[1]{\ensuremath{\boldsymbol{#1}}}
\newcommand{\be}{\begin{equation}}
\newcommand{\ee}{\end{equation}}
\newcommand{\ba}{\begin{array}}
\newcommand{\ea}{\end{array}}
\newcommand{\bea}{\begin{eqnarray}}
\newcommand{\eea}{\end{eqnarray}}
\newcommand{\bean}{\begin{eqnarray*}}
\newcommand{\eean}{\end{eqnarray*}}

\definecolor{white}{rgb}{1,1,1}

\newtheorem{theorem}{Theorem}
\newtheorem{lemma}{Lemma}
\newtheorem{example}{Example}

\newtheorem{property}{Property}
\newtheorem{definition}{Definition}
\newtheorem{corollary}{Corollary}

\begin{document}

\title{An Information Theoretic Charachterization of Channel Shortening Receivers}

\author{Fredrik~Rusek, and~Ove Edfors,~\IEEEmembership{Member,~IEEE}\\
Department of Electrical and Information Technology, Lund University, Sweden\\
\{fredrik.rusek,ove.edfors\}@eit.lth.se
\thanks{The work of F.~Rusek was supported by SSF through the distributed antenna project. This paper was presented in part at the 2013 Asilomar Conference on Signals, Systems \& Computers.
}
}

\maketitle 

\begin{abstract}
Optimal data detection of data transmitted over a linear channel can always be implemented through the Viterbi algorithm (VA). However, in many cases of interest the memory of the channel prohibits application of the VA. A popular and conceptually simple method in this case, studied since the early 70s, is to first filter the received signal in order to shorten the memory of the channel, and then to apply a VA that operates with the shorter memory. We shall refer to this as a channel shortening (CS) receiver. Although studied for almost four decades, an information theoretic understanding of what such a simple receiver solution is actually doing is not available.

In this paper we will show that an optimized CS receiver is implementing the chain rule of mutual information, but only up to the shortened memory that the receiver is operating with. Further, we will show that the tools for analyzing the ensuing achievable rates from an optimized CS receiver are precisely the same as those used for analyzing the achievable rates of a minimum mean square error (MMSE) receiver. 
\end{abstract}


\section{Introduction}
In 1972, Forney \cite{F72} showed that the Viterbi Algorithm (VA) can be applied to intersymbol interference (ISI) channels in order to implement maximum likelihood (ML) detection. However, the complexity of the VA is exponential in the memory of the channel which prohibits its use in many cases of interest. As a remedy, Falconer and Magee proposed in 1973 the concept of channel shortening \cite{FM73}, also known as combined linear and Viterbi equalization. The concept is straightforward: (i) Filter the received signal with a channel shortening filter so that the effective channel has much shorter duration than the original channel, (ii) Apply the VA to the shorter effective channel. 
Although Falconer and Magee's original paper dealt solely with ISI, the concept extends straighforwardly to general linear channels in which case "filter with a channel shortening filter" should be interpreted as a matrix multiplication. After a QR/QL factorization, the VA is then applied. Albeit CS is conceptually simple, the achievable rates that can be supported by such receiver was first derived as late as 2012 in \cite{RP12}. A derivation for the case of ISI was available already 2000 in \cite{AL00}. However, the system model in \cite{AL00} is limited and cannot reach the same results as \cite{RP12}. We shall come back to \cite{AL00} later in the paper when sufficient notation has been introduced so that the drawbacks of \cite{AL00} can be better illuminated. While \cite{RP12} established the optimal parameters for the CS receiver, no insights into the nature of the optimized CS receiver was given. In this paper we analyze the optimal CS receiver from an information theoretic perspective. The two main findings are: (i) An optimized CS receiver implements the chain rule of mutual information up to the reduced memory of the receiver and (ii) The tools for analyzing the achievable rates of CS are precisely the same as those used for analyzing the rates of MMSE receivers \cite{MCT}.

\section{System model}
We consider a received signal that can be described by means of the following discrete-time model
\be \label{sysmod}
\vec{y}=\vec{Hx}+\vec{n},
\ee
where $\vec{y}$ is the $\nr\times 1$ received vector, $\vec{H}$ an arbitrary channel matrix of dimension $\nr \times \nt$ that is perfectly known to the receiver, $\vec{x}$  an $\nt\times 1$ vector comprising the transmitted symbols drawn from an alphabet  $\mathcal{A}$,  and $\vec{n}$  an $\nr\times 1$ noise vector. We assume that $\vec{x}$ is distributed as a zero-mean circularily-symmetric complex Gaussian distributed vector with covariance $\vec{I}_{\nt}$ and that $\vec{n}$ is distributed as a zero-mean circularily-symmetric complex Gaussian distributed vector with covariance $N_0\vec{I}_{\nr}$. Note that we are not imposing any structure upon the matrix $\vec{H}$, so that  (\ref{sysmod}) encompasses many communication systems, such as multiple-input multiple-output (MIMO), intersymbol interference (ISI), MIMO-ISI channels, intercarrier interference (ICI) etc. In our subsequent analysis the underlying structure of the channel matrix is irrelevant - the same results apply in all cases - but we point out that the Toeplitz structure of the matrix for the ISI cases can be used to simplify the formulas.

An optimal receiver operates on the basis of the conditional probability density function (pdf)
\be \label{optdet}
p_{\vec{Y}|\vec{X}}(\vec{y}|\vec{x})\propto \exp\left(-\frac{\|\vec{y}-\vec{Hx}\|^2}{N_0}\right) 
\ee
where we use bold upper case letters for random vectors and bold lower case letters for their realizations. Matrices are always denoted by bold upper case letters no matter whether they are deterministic or random. The optimal receiver  can reach the information rate of the channel 
\bea I_{\mathrm{R}} &=& I(\vec{Y};\vec{X}) \nonumber \\
&=& \log\left(\det\left(\vec{I}_{\nt}+\vec{H}^{\mathrm{H}}\vec{H}\right)\right). \eea
In the case of ISI channels, limits and normalization must be included,
\be I_{\mathrm{R}} = \lim_{\nt \to \infty} \frac{1}{\nt} I(\vec{Y};\vec{X}). \ee

The Gaussian assumption on the inputs is made in order to reach mutual information results, but Gaussian inputs are impractical and finite cardinality inputs are used in practice. However, Gaussian inputs are still most relevant in communication theory as they represent very well the rates that can be achievad with, e.g., quadrature amplitude modulation (QAM) constellations.  With QAM, the optimal receiver can be implemented over a trellis with memory $L$. The memory $L$ is  determined by the channel matrix $\vec{H}$ and is defined formally next.
\begin{definition} Define $\vec{G}=\vec{H}^{\mathrm{H}}\vec{H}$. The memory of $\vec{H}$ is the smallest number $L$ that satisfies
$$G_{k,\ell}=0, \forall|k-\ell|>L.$$ 
\end{definition}
Commonly, the receiver iterates between decoding the outer error correcting code and detection of the data symbols. In that case, the VA is replaced by the BCJR algorithm which operates over the same trellis. With  iterative receivers, achievability of the rate $I(\vec{Y};\vec{X})$ (computed for the finite cardinality alphabet $\mathcal{A}$) is not guaranteed. However, even for an iterative receiver, the rate $I(\vec{Y};\vec{X})$ is of significant operational meaning in the sense that iterative receivers can operate close the information rate of the channel if the overall transceiver system is properly designed.

For MIMO and ICI, the memory is typically "full" in the sense that $L=\nt-1$. In those cases, the optimal detector is operating over a tree rather than over a trellis. Nevetheless, after linear filtering, we shall compress the memory of the channel, so that trellis processing can be applied. By applying a suitable permutation of the columns of $\vec{H}$ one may obtain a smaller memory. This paper does, however, not cover such permutations.

\subsection{Classical CS}\label{ccs}
Since the number of states $|\mathcal{A}|^L$ can be very large in practice, it is of interest to seek sub-optimal receivers that reduces the number of states in the trellis. In the case of a "`full"' memory, i.e., $L=\nt-1$, it is of interest to convert the tree into a much smaller trellis. Falconer and Magee's classical CS proceeds via the following steps,
\begin{enumerate}
\item Filter the signal $\vec{y}$ with a matrix $\vec{W}$, to obtain $\vec{r} =\vec{Wy}$. 
\item Impose the structure $\vec{r} = \vec{Fx}+\vec{w}$, where $\vec{F}$ is a memory $K<L$ matrix and $\vec{w}$ a noise vector. 
\item Further process the signal $\vec{r}$ as if $\vec{F}$ is the true channel and $\vec{w}$ is white noise. 
\end{enumerate}
In terms of a conditional pdf, classical CS can be expressed as if the receiver is operating on the basis of the mismatched function
\be \label{CSdet}
T(\vec{y}|\vec{x})=\exp\left(-\|\vec{Wy}-\vec{Fx}\|^2\right).
\ee
We point out that $T(\vec{y}|\vec{x})$ does not qualify as a conditional pdf as it does not in general satisfy $\int T(\vec{y}|\vec{x})\mathrm{d}\vec{y}=1$, but this is irrelevant. Further, it is no loss of generality to assume a unit noise density, as the two matrices $\vec{W}$ and $\vec{F}$ can be scaled at will.

\subsection{A new framework for CS}\label{frm}
To the best of the authors' knowledge, all previous papers dealing with CS detection has been based on the model (\ref{CSdet}) and the goal has been to optimize the receiver parameters $\vec{W}$ and $\vec{F}$. However, the system model (\ref{CSdet}) is not the only system model for CS, and in fact not even the most suitable. Neglecting factors that do not depend on $\vec{x}$, the receiver function $T(\vec{y}|\vec{x})$ can be expressed as
\be \label{CSrew} T(\vec{y}|\vec{x})\propto \exp\left(2\mathcal{R}\left\{\vec{x}^{\mathrm{H}}\vec{F}^{\mathrm{H}}\vec{Wy}\right\}-\vec{x}^{\mathrm{H}}\vec{F}^{\mathrm{H}}\vec{Fx}\right).
\ee
The receiver can work directly with (\ref{CSrew}), with no increase in computational complexity compared to a receiver that work with the model (\ref{CSdet}). A VA operating with (\ref{CSrew}) was first proposed by Ungerboeck in 1974 \cite{U74} and its BCJR-version by Colavolpe and Barbieri in \cite{CB05}. The model (\ref{CSrew}) is commonly refered to as the Ungerboeck model. 

We now modify $T(\vec{y}|\vec{x})$ in order to obtain an alternative framework for CS. In \cite{RP12} it was proposed to abandon (\ref{CSdet}) in favor of 
\be \label{CSubm} \tilde{T}(\vec{y}|\vec{x})= \exp\left(2\mathcal{R}\left\{\vec{x}^{\mathrm{H}}\vec{H}_{\mathrm{r}}\vec{y}\right\}-\vec{x}^{\mathrm{H}}\vec{G}_{\mathrm{r}}\vec{x}\right),
\ee
where $\vec{H}_{\mathrm{r}}$ is an arbitrary $\nt\times \nr$ matrix and $\vec{G}_{\mathrm{r}}$ is a Hermitian $\nt \times \nt$ matrix where only the elements along the center $2K+1$ diagonals can take non-zero values. 
Again, based on \cite{U74} the receiver can be implemented also for $\tilde{T}(\vec{y}|\vec{x})$ and leads to  memory $K$ of the receiver. Altogether, from a conceptual and a computational complexity point of view, it is irrelevant whether the receiver is implemented over $T(\vec{y}|\vec{x})$ or $\tilde{T}(\vec{y}|\vec{x})$, but as we discuss next, the latter function offers an advantage over the former.

\begin{property}
The receiver function $\tilde{T}(\vec{y}|\vec{x})$ specifies a more general mismatched receiver framework for CS than $T(\vec{y}|\vec{x})$ as the matrix $\vec{G}_{\mathrm{r}}$ need not be positive semi-definite as the matrix $\vec{F}^{\mathrm{H}}\vec{F}$ must be. For a given memory $K$, the complexity is identical in both cases.
\end{property}

Ostensibly, it may appear as if Property 1 lacks operational interest as one is tempted to assume that an optimized system would use an indefinite $\vec{G}_{\mathrm{r}}$ only in very rare special cases, but this is not the case. Whenever the channel matrix $\vec{H}$ contains one or more small, but still strictly positive, eigenvalues, the optimal matrix $\vec{G}_{\mathrm{r}}$  is often indefinite. We provide some numerical examples of this in Section \ref{comp}.

If we set $K=\nt-1$ the mismatched receiver function $T(\vec{y}|\vec{x})$ can be made proportional to the true conditional pdf, which means that the optimal receiver is included as a special case of CS. Further, with $K=0$ we can reach the linear MMSE equalizer. Hence, CS has these two well known receivers as limiting cases and, as we will show later, CS shares many properties with the MMSE equalizer.

For the classical CS model (\ref{CSdet}) and ISI channels, it was shown in \cite{AL00} how to optimize $\vec{W}$ and $\vec{F}$. However, the optimization done in \cite{AL00} was in fact done over $\Gr$, and then it was concluded that $\vec{F}$ could be recaptured from $\Gr$. However, the found $\Gr$ in \cite{AL00} is not positive semi-definite in general, so it is not possible to recapture $\vec{F}$. Altogether, the method in \cite{AL00} is based on (\ref{CSdet}), does not adress general linear channels, and its optimization method fails in general.

\subsection{A Special case of CS} \label{bdcs}
A popular special case of CS is to use a block diagonal form for $\vec{G}_{\mathrm{r}}$ \cite{SUMIS}. We assume that $\Gr$ contains $M$ blocks of dimensions $K_m\times K_m,\; 1\leq m\leq M,$ along the main diagonal, with 
$$\sum_{m=1}^M K_m =\nt.$$
The rationale of this simplification is that the detection is broken up into $M$ trees of depths $K_m$, rather than performing the detection over a single trellis of memory $K$ as in normal CS. Since detection complexity is largely determined by the largest value of $K_m$, all blocks should preferably have the same dimension $K_m=K$, but this is not  possible for all parameter combinations. An important property of such scheme is 
\begin{property}
With a block diagonal constraint on $\Gr$, the optimal $\Gr$ is always positive semi-definite.
\end{property}
We point out that "optimal" is with respect to generalized mutual information, which will be made more precise in Section \ref{rates}. This implies that for a block diagonal $\Gr$ there is no gain in using the new framework from Section \ref{frm} as $\tilde{T}(\vec{y}|\vec{x})$ can always be cast in the form of $T(\vec{y}|\vec{x})$ from Section \ref{ccs}. 

\section{Analysis of the Achievable rates of CS} \label{rates}
With a receiver that operates with $\tilde{T}(\vec{y}|\vec{x})$ instead of the true $p_{\vec{Y}|\vec{X}}(\vec{y}|\vec{x})$, the information rate of the channel
$I_{\mathrm{R}}$ cannot be reached in general.  Instead, the relevant performance measure is the generalized mutual information (GMI). The GMI establishes a lower bound on the achievable rate that can be supported with the mismatched receiver function. For given receiver parameters $\Hr$ and $\Gr$, the GMI, in nats/channel input, equals
$$I_{\mathrm{GMI}}(\Hr,\Gr)=-\mathbb{E}\left[\log\left(\tilde{T}(\vec{y})\right)\right]+\mathbb{E}\left[\log\left(\tilde{T}(\vec{y}|\vec{x})\right)\right],$$
where
$$\tilde{T}(\vec{y})=\int \tilde{T}(\vec{y}|\vec{x})p_{\vec{X}}(\vec{x})\mathrm{d}\vec{x},$$
and the expectations are with respect to the true conditional pdf  $p_{\vec{Y}|\vec{X}}(\vec{y}|\vec{x})$. 

Maximization of $I_{\mathrm{GMI}}$ over the two matrices $\Hr$ and $\Gr$ was carried out in \cite{RP12}. However, the expression for the optimal $\Gr$ given in \cite{RP12} is complicated, and in our next result we give an alternative formulation of $\Gr$. In order to do so, we introduce notation from \cite{KM00}.
\begin{definition}[From \cite{KM00}] \label{deflband}
The matrix $\vec{R}$ is called the $L$-band extension of an $N\times N$ matrix $\vec{C}$ if its inverse $\vec{R}^{-1}$ is related to $\vec{C}$ as 
\be \label{lband} \vec{R}^{-1} = \sum_{n=1}^{N-L-1} \mathcal{\vec{P}}_{n+L}^n\left[\left(\vec{C}_{n+L}^n\right)^{-1}\right]-\sum_{n=2}^{N-L} \mathcal{\vec{P}}_{n+L-1}^n\left[\left(\vec{C}_{n+L-1}^n\right)^{-1}\right]+\sum_{n=2}^{N-L} \mathcal{\vec{P}}_{n+L}^n\left[\left(\vec{C}_{n+L}^n\right)^{-1}\right],\ee
where $\mathcal{\vec{P}}_j^i[\vec{X}]$ is an $N\times N$ matrix whose principal submatrix spanning columns (and rows) $i$ through $j$ is equal to $\vec{X}$, and all other elements are zero. The matrix $\vec{C}_{n+\ell}^n$ is the $(\ell+1) \times (\ell+1)$ principal submatrix spanning columns (and rows) $n$ through $n+\ell$ of $\vec{C}$. 
For graphical visualization of (\ref{lband}), we refer the reader to \cite[Eqs. (5) and (7)]{KM00}. 
\end{definition}

We can now give an alternative (to the one in \cite{RP12}) formulation of the optimal $\Gr$ in terms of Definition \ref{deflband}. For completeness, we also state the optimal $\Hr$.
\begin{theorem} \label{thm1}
The solution to 
$$\Hr^{\mathrm{opt}}=\arg \max_{\Hr}I_{\mathrm{GMI}}(\Hr,\Gr)$$
for any $\Gr$ is given by
$$\Hr^{\mathrm{opt}} =(\vec{I}_{\nt}+\Gr)\vec{H}^{\mathrm{H}}\left[\vec{H}\vec{H}^{\mathrm{H}}+N_0\vec{I}_{\nr}\right]^{-1}.$$
For the optimal $\Hr$, it holds that the optimal $\Gr$, i.e., 
$$\Gr^{\mathrm{opt}}=\arg \max_{\Gr}I_{\mathrm{GMI}}(\Hr^{\mathrm{opt}},\Gr),$$
is such that $(\vec{I}_{\nt}+\Gr^{\mathrm{opt}})^{-1}$ is the $K$-band extension of the MMSE matrix
$$\vec{B}=\vec{I}_{\nt}-\vec{H}^{\mathrm{H}}(\vec{H}\vec{H}^{\mathrm{H}}+\vec{I}_{\nr})^{-1} \vec{H}.$$
In other words,
\be \label{lband2}  \vec{I}_{\nt}+\Gr^{\mathrm{opt}}= \sum_{n=1}^{\nt-K-1} \mathcal{\vec{P}}_{n+K}^n\left[\left(\vec{B}_{n+K}^n\right)^{-1}\right]-\sum_{n=2}^{\nt-K} \mathcal{\vec{P}}_{n+K-1}^n\left[\left(\vec{B}_{n+K-1}^n\right)^{-1}\right]+\sum_{n=2}^{\nt-K} \mathcal{\vec{P}}_{n+K}^n\left[\left(\vec{B}_{n+K}^n\right)^{-1}\right].\ee
\end{theorem}
The proof is given in Appendix A.

An immediate corollary that sheds further light on the structure of $\Gr^{\mathrm{opt}}$ is
\begin{corollary} \label{inversechar}
Let $\mathrm{diag}_K(\vec{X})$ be a matrix of equal dimensions as $\vec{X}$ that equals $\vec{X}$ along the center $2K+1$ diagonals and is zero elsewhere, i.e., if $\vec{Z}=\mathrm{diag}_K(\vec{X})$ then
$$Z_{k\ell}=\left\{\begin{array}{ll} X_{k\ell},& |k-\ell|\leq K \\ 0, &|k-\ell|> K. \end{array} \right.$$
The optimal $\Gr^{\mathrm{opt}}$ satisfies
$$\mathrm{diag}_K\left([\vec{I}_{\nt}+\Gr^{\mathrm{opt}}]^{-1}\right)=\mathrm{diag}_K\left(\vec{B}\right).$$
\end{corollary}
The proof is given in Appendix B. The proof makes use of \cite[Theorem 2]{KM00}. However, in the light of Theorem \ref{thm1} it is possible to sharpen \cite[Theorem 2]{KM00} and, although we shall not make use of the sharpened result, we take the opportunity to do this in Appendix C.

While Theorem \ref{thm1} and its Corollary \ref{inversechar} dealt with the structure of the optimal receiver parameters, nothing was said about the rate $I_{\mathrm{GMI}}$. We next turn our attention to such result, with a corollary that simplifies the expression for the GMI.
\begin{corollary} \label{corsimple}
For the optimal $\Hr^{\mathrm{opt}}$ and $\Gr^{\mathrm{opt}}$, we have
$$I_{\mathrm{GMI}}(\Hr^{\mathrm{opt}},\Gr^{\mathrm{opt}})=\log\left(\det\left(\vec{I}_{\nt}+\Gr^{\mathrm{opt}}\right)\right).$$
\end{corollary}
The proof is given in Appendix D.

Continuing the charachterization of the optimal $I_{\mathrm{GMI}}$, we next give our main theorem.
\begin{theorem} \label{main}
Let $\bar{\vec{H}}_{[k,n]}$ be the same matrix as $\vec{H}$ but with columns  $[k,k+1,\ldots,n]$ removed. Let $\vec{G}_{[k,n]}=\bar{\vec{H}}_{[k,n]}^{\mathrm{H}}\bar{\vec{H}}_{[k,n]}$ and use the convention $\vec{G}_{[k,n]}=\vec{G}=\vec{H}^{\mathrm{H}}\vec{H}$ for $n<k$. For  $\Hr^{\mathrm{opt}}$ and $\Gr^{\mathrm{opt}}$, we have
\bea I_{\mathrm{GMI}}(\Hr^{\mathrm{opt}},\Gr^{\mathrm{opt}})&=&\log\left(\det\left(\vec{I}_{\nt}+\frac{\vec{G}}{N_0}\right)\right)\nonumber \\
&&-\sum_{k=1}^{\nt-K}\log\left(\det\left(\vec{I}_{\nt}+\frac{\vec{G}_{[k,k+K]}}{N_0}\right)\right)\nonumber \\
&&+\sum_{k=2}^{\nt-K}\log\left(\det\left(\vec{I}_{\nt}+\frac{\vec{G}_{[k,k+K-1]}}{N_0}\right)\right).
\eea
\end{theorem}
The proof is given in Appendix E. From this point and onwards we shall assume that the two matrices $\Hr$ and $\Gr$ are always optimized and we shall therefore drop the superscript $\mathrm{opt}$. We will also use the shorthand notation $I_{\mathrm{GMI}}$ instead of $I_{\mathrm{GMI}}(\Hr^{\mathrm{opt}},\Gr^{\mathrm{opt}})$.

In \cite{MCT} the MMSE equalizer was analyzed and the following formula for the achievable rate was established,
\bea\label{mmse}
I_{\mathrm{MMSE}}&=&\nt\log\left(\det\left(\vec{I}_{\nt}+\frac{\vec{G}}{N_0}\right)\right)\nonumber \\
&&-\sum_{k=1}^{\nt}\log\left(\det\left(\vec{I}_{\nt}+\frac{\vec{G}_{[k,k]}}{N_0}\right)\right).
\eea
By inspection, it can be seen that by setting $K=0$, Theorem \ref{main} collapses into $I_{\mathrm{MMSE}}$ in (\ref{mmse}). The structure of the formula for $I_{\mathrm{GMI}}$ of CS detection is closely related to that of
$I_{\mathrm{MMSE}}$. With MMSE detection, single columns are removed from  $\vec{H}$ which produce matrices $\vec{G}_{[k,k]}$. With CS, $K+1$ columns are removed. With random MIMO channels where all columns of the matrix are independent and identically distributed (IID), an analysis of the effect of removing $K+1$ columns from an $\nr\times \nt$ matrix is the same as an analysis of the effect of removing a single column from an $\nr\times (\nt-K)$  matrix. Thus,  an analysis of $I_{\mathrm{GMI}}$ of CS detection for $\nr\times \nt$ MIMO is equivalent to an analysis of the achievable rates of MMSE for $\nr\times (\nt-K)$ MIMO. There is an abundance of literature dealing with analysis of the MMSE receiver, and essentially all of those results can be carried over to CS detection through Theorem \ref{main}. We will examplify this in Section \ref{numres}. 

We close this section with a re-formulation of Theorem \ref{main} that sheds further light of the nature of CS detection. Recall that by using the chain rule of mutual information, the information rate of the channel can be expressed as
$$I_{\mathrm{R}}=\sum_{k=1}^{\nt} I(\vec{Y};X_k|X_{k-1},\ldots,X_1).$$
We have
\begin{corollary} \label{chain}
The rate $I_{\mathrm{GMI}}$ in Theorem \ref{main} can be expressed as
$$I_{\mathrm{GMI}}=\sum_{k=1}^{\nt} I(\vec{Y};X_k|X_{k-1},\ldots,X_{k-K}).$$
\end{corollary}
The proof is given in Appendix F.

Corollary \ref{chain} is most intuative: A properly optimized CS detector based on (\ref{CSubm}) implements the chain rule of mutual information, but only up to the reduced memory of the receiver.

In order to compare the ensuing rate from a banded $\Gr$ with that of a block diagonal $\Gr$, let us formally state the latter rate in
\begin{lemma} \label{block}
With a block diagonal structure of $\Gr$ with $M$ blocks, each one of dimension $K_m\times K_m$, we have
$$I_{\mathrm{GMI}}=\sum_{m=1}^M\sum_{k=1}^{K_m}I\left(\vec{Y};X_{T_m+k}|X_{T_m+k-1},...,X_{T_m}\right),$$
where $T_m=\sum_{\ell=1}^{m-1}K_{\ell}.$
\end{lemma}
The proof is given in Appendix G. The meaning of Lemma \ref{block} is that the chain rule of mutual information is implemented, but conditioning does not carry over across the blocks.

\section{A Comparison between the two Models for CS} \label{comp}
Let $\lambda_{\min}$ denote the smallest eigenvalue of the optimal $\Gr$. We know that whenever $\lambda_{\min}<0$, no factorization $\vec{F}^{\mathrm{H}}\vec{F}=\Gr$ is possible which means that a CS detector based on (\ref{CSdet}) cannot reach the optimal solution for the Ungerboeck based model (\ref{CSubm}). Further, while the optimal $\Gr$ has a closed form solution, we have not been able to find a closed form solution for the optimal $\vec{F}$ to use in (\ref{CSdet}). Clearly, whenever $\lambda_{\min}\geq 0$, the optimal $\vec{F}$ is the Cholesky factorization of the optimal $\Gr$, but whenever $\lambda_{\min}<0$, it is unclear how to solve for the optimal $\vec{F}$. In this section we shall provide a numerical optimization method to find such optimal $\vec{F}$ and evaluate how sub-optimal it is through simulations.

The optimization problem to solve is
$$\vec{F}^{\mathrm{opt}}=\arg\max_{\vec{F}} f\left(\vec{F},\vec{B}\right)$$
where $\vec{F}$ is an upper triangular matrix that only takes non-zero values along the first $K+1$ diagonals and where, from \cite{RP12},
$$f\left(\vec{F},\vec{B}\right)\triangleq\log\left(\det\left(\vec{I}_{\nt}+\vec{F}^{\mathrm{H}}\vec{F}\right)\right)-\mathrm{Tr}\left(\left(\vec{I}_{\nt}+\vec{F}^{\mathrm{H}}\vec{F}\right)\vec{B}\right)+\nt.$$
The function $f\left(\vec{F},\vec{B}\right)$ is a concave function of $\vec{F}$. Since the constraints on $\vec{F}$ are linear, we know that any local maximum is the global maximum. 

The gradient with respect to the matrix $\vec{F}$ is
$$\nabla_{\vec{F}}f\left(\vec{F},\vec{B}\right) = 2\vec{F}\left(\vec{I}_{\nt}+\vec{F}^{\mathrm{H}}\vec{F}\right)^{-1}-2\vec{F}\vec{B}.$$
This gradient is computed at all positions in the matrix $\vec{F}$, not only the ones that are allowed to take non-zero values. We therefore introduce the special notation $\mathrm{diag}_K^{\mathrm{Up}}(\vec{X})$ to denote a matrix of the same size as $\vec{X}$, with identical values on the first $K+1$ upper diagonals and zeros elsewhere. That is, if $\vec{Z}=\mathrm{diag}_K^{\mathrm{Up}}(\vec{X})$, then
$$Z_{k\ell}=\left\{\begin{array}{ll} X_{k\ell}, & k\leq \ell \leq k+K \\0,& \mathrm{otherwise}.\end{array}\right.$$

In the case $\lambda_{\min}<0$ we propose an iterative optimization procedure. In the first step we use $\vec{F}_{(0)}$, the Cholesky factorization of a regularized version of $\Gr$, 
$$\vec{F}_{(0)}^{\mathrm{H}}\vec{F}_{(0)}=\Gr-\lambda_{\min}\vec{I}_{\nt},$$
as initialization. We then proceed in the direction of the gradient so that in the $i$th iteration, we construct
\bea \label{iterative} \vec{F}_{(i)}&=&\vec{F}_{(i-1)}+\mathrm{diag}_K^{\mathrm{Up}}\left(\nabla_{\vec{F}}f\left(\vec{F}_{(i-1)},\vec{B}\right) \right)\nonumber \\
&=&\vec{F}_{(i-1)}+\mathrm{diag}_K^{\mathrm{Up}}\left(2\vec{F}_{(i-1)}\left(\vec{I}_{\nt}+\vec{F}^{\mathrm{H}}_{(i-1)}\vec{F}_{(i-1)}\right)^{-1}-2\vec{F}_{(i-1)}\vec{B}\right).\eea
We iterate this procedure until, e.g., the maximum element of the $\mathrm{diag}_K^{\mathrm{Up}}\left(\nabla_{\vec{F}}f\left(\vec{F}_{(i-1)},\vec{B}\right) \right)$ is smaller than some pre-selected threshold $\epsilon$. Based on tests, the iterative optimization is highly stable and converges to the global maximum within a few iterations.

\begin{figure}[t]
\begin{center}
\scalebox{1}{\hspace*{-0mm}\includegraphics*{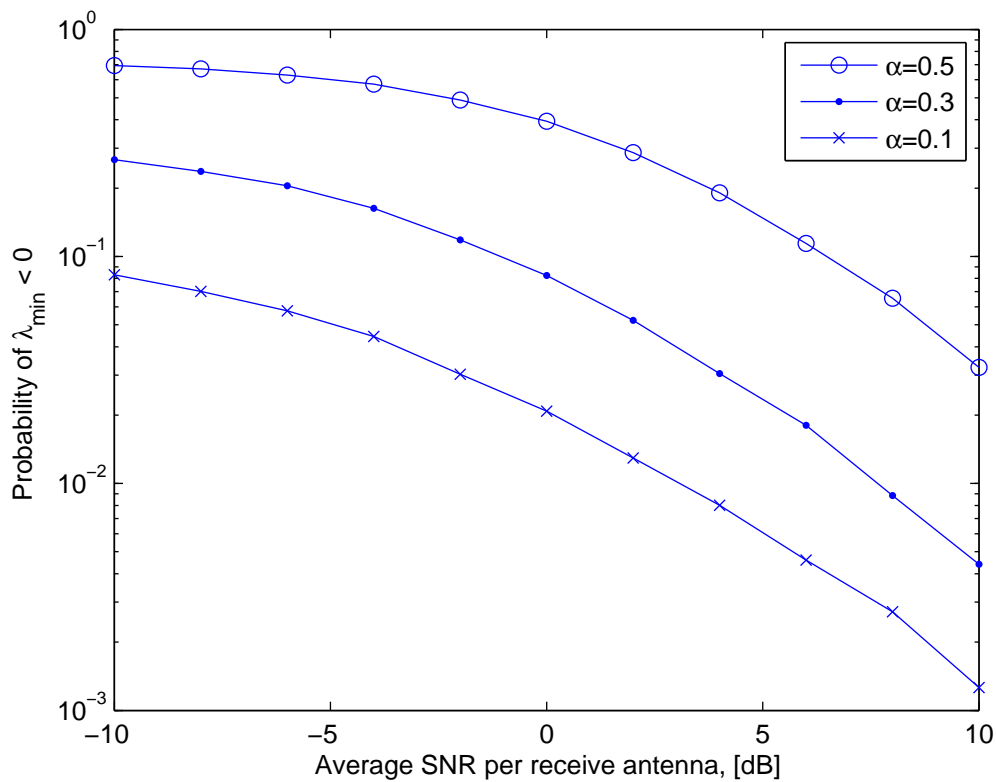}}
\vspace*{-3mm}
\caption{\label{fig1} Probability that $\lambda_{\min}<0$ for $5\times 5$ MIMO channels with a Kronecker correlation model.  }
\end{center}
\vspace*{-7mm}  
\end{figure}

\begin{figure}[t]
\begin{center}
\scalebox{1}{\hspace*{-0mm}\includegraphics*{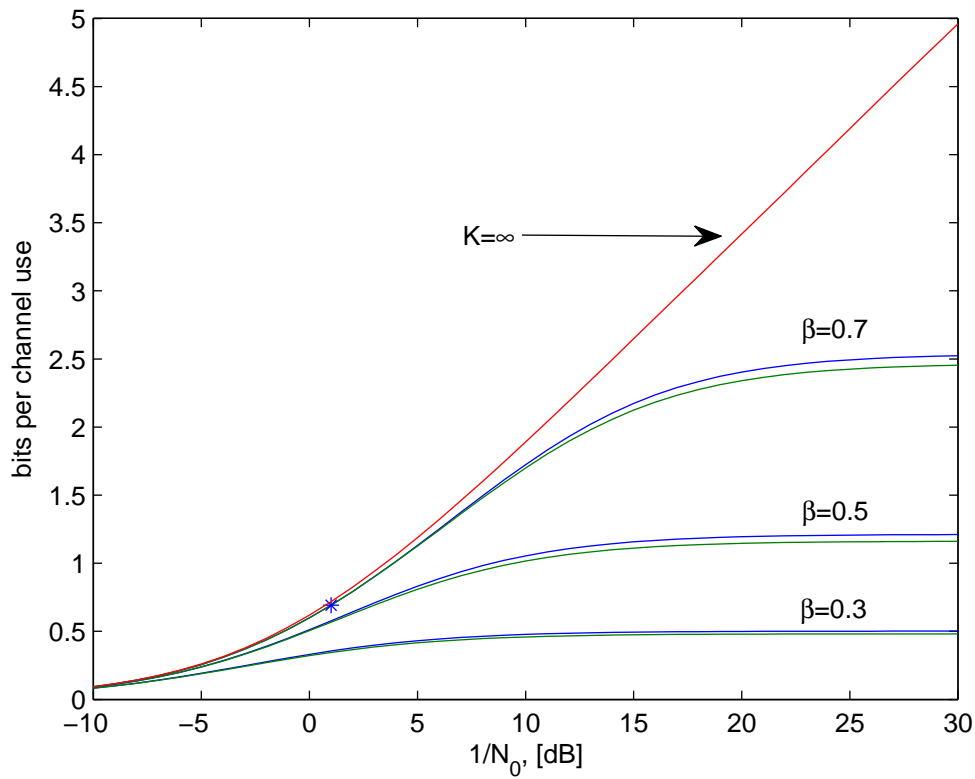}}
\vspace*{-3mm}
\caption{\label{fig2} Achievable rates for ISI channels with transfer functions (\ref{isic}). Within each set of curves, the upper curve shows the model (\ref{CSrew}) while the lower shows the classical model (\ref{CSdet}). In all cases $K=1$ except for the curve marked with "`$K=\infty$"' which is computed for $\beta=0.7$.}
\end{center}
\vspace*{-7mm}  
\end{figure}

We next turn to numerical results in order to quantify how sub-optimal the classical framweork for CS detection is. We consider $5\times 5$ MIMO channels that are correlated according to a Kronecker correlation model. Both the rows are and the columns of the channel matrix are correlated according to a Toeplitz matrix 
$$\vec{\Phi}=\mathrm{Toeplitz}[1\,\alpha\,\alpha^2\,\alpha^3\,\alpha^4].$$
We have chosen $\alpha\in\{0.1,0.3,0.5\}$ which represent "`low"', "`medium"', and "`high"' correlation according to the 3GPP test cases. If one compares the ensuing achievable rates from an optimized model (\ref{CSdet}) with those from an optimized Ungerboeck model (\ref{CSubm}), the results are virtually indistinguishable. Typically, the rate with the classical model is around 99.95\% of the Ungerboeck rate in the cases when $\lambda_{\min}<0$. In fact, already the regularized $\vec{F}_{(0)}$ has performance around 99.9\% of the optimal rate. Hence, we omit to show any plots as all rate curves are anyway overlapping. What is more interesting is to note how often it happens that $\lambda_{\min}<0$. Whenever this happens, an implementation of a detector based on (\ref{CSdet}) must first regularize $\Gr$ before taking the Cholesky factorization.
In Figure \ref{fig1} we show the probability of $\lambda_{\min}<0$ for $5\times 5$ MIMO channels. 
As can be seen, at low SNR, it frequently happens that $\lambda_{\min}<0$, so that this cannot be ignored in an implementation. We can also see that the $\lambda_{\min}<0$ is much more frequently occuring whenever the correlation is strong. For this reason we consider ISI channels that are charachterized by perfectly bandlimited low pass filters; the columns of the resulting matrix $\vec{H}$ are close to parallel. Let the transfer function of a time discrete impulse response be 
\be \label{isic} |H(\omega)|^2=\left\{\begin{array}{ll} \frac{1}{\beta}, & |\omega|\leq \beta \pi \\0,& |\omega|> \beta \pi. \end{array} \right.\ee
The ISI case is a special case of the system model (\ref{sysmod}) so that the same techniques can be applied to optimize the receiver parameters. In Figure \ref{fig2} we show the achievable rates of the two optimized models. We give results for $K=1$ with $\beta\in\{0.3,0.5,0.7\}$ which means that the impulse responese represent strong narrowband channels.
Within each set of curves, the upper one is the model (\ref{CSubm}) while the lower is the classical model (\ref{CSdet}). The curve marked with $K=\infty$ shows the rate for a full complexity detector with $\beta=0.7$. The asterisk shows the location where $\lambda_{\min}<0$ for the first time, i.e., to the left of the asterisk there is no difference between the two models, while a difference is observed to the right of it. For $\beta=0.3$ and 0.5, these locations are below -10 dB, and are not shown.
As can be seen, there is a small performance difference at high SNR. Also, in all cases, the achievable rates saturate at high SNR. We will get back to the reason for this in Section \ref{numres}.

\section{Applications} \label{numres}
Let us now consider the ergodic achievable rates of $\nr\times \nt$ MIMO channels comprising IID complex Gaussian random variates, each one with zero mean and unit variance, with CS detection.
Since we are interested in ergodic rates and the channel elements are IID, the formula for $\mathbb{E}[I_{\mathrm{GMI}}]$ simplifies. This is so since $\vec{G}_{[k,k+n]}$ is statistically equivalent to $\vec{G}_{[k+p,k+n+p]}$ for any $p$. 
Let us introduce the notation
$$\bar{I}[\nt,\nr,\mathrm{snr}]\triangleq \mathbb{E} \left[\log\left(\det\left(\vec{I}_{\nt}+\frac{\mathrm{snr}}{\nt}\vec{H}^{\mathrm{H}}\vec{H}\right)\right)\right]$$
where $\vec{H}$ is an IID complex Gaussian random matrix of dimension $\nr\times \nt$.
Then Theorem \ref{main} gives
\bea \label{eqw} \mathbb{E}[I_{\mathrm{GMI}}(\mathrm{snr})]&=&\bar{I}[\nt,\nr,\mathrm{snr}]\!-\!(\nt\!-\!K)\bar{I}[\nt\!-\!K\!-\!1,\nr,\mathrm{snr}]\nonumber \\
&&+(\nt-K-1)\bar{I}[\nt-K,\nr,\mathrm{snr}].
\eea
Let us now consider the high signal-to-noise-ratio (SNR) expansion of the achievable ergodic rate 
$$\mathbb{E}[I_{\mathrm{GMI}}(\mathrm{snr})]=S_{\infty}(\log(\mathrm{snr})-L_{\infty}),$$
where $S_{\infty}$ is the high SNR asymptotic slope 
$$S_{\infty}=\lim_{\mathrm{snr}\to \infty} \frac{\mathbb{E}[I_{\mathrm{GMI}}(\mathrm{snr})]}{\log(\mathrm{snr})}$$
and $L_{\infty}$ is the high SNR power offset given by
\be \label{linf} L_{\infty}=\lim_{\mathrm{snr}\to \infty}\left(\log(\mathrm{snr})-\frac{\mathbb{E}[I_{\mathrm{GMI}}(\mathrm{snr})]}{S_{\infty}}\right).\ee

For full complexity detection, a well known result is \cite{MCT}
\be \label{slopefull} S_{\infty}^{\mathrm{full}}=\lim_{\mathrm{snr}\to \infty}\frac{\bar{I}[\nt,\nr,\mathrm{snr}]}{\log(\mathrm{snr})}=\min(\nr,\nt).\ee
For the MMSE equalizer, i.e., a CS detector with $K=0$, a few manipulations gives
$$S_{\infty}^{\mathrm{MMSE}}=\left\{\begin{array}{ll}\nt, & \nr\geq \nt \\ 0,& \nr<\nt.\end{array}\right.$$
Thus, for MMSE equalization, the asymptotic slope of the ergodic rate is zero if the number of receive antennas is less than the number of transmit antennas. As we shall see next, CS can compensate for the lack of receive antennas.
\begin{lemma} \label{cslemma}
For an optimized CS detector with memory $K$ we have
$$S_{\infty}^{\mathrm{CS}}=\left\{\begin{array}{ll}\nt, & \nr\geq \nt \\ \nr,& \nr<\nt, \nr+K\geq \nt \\ 0,& \mathrm{otherwise}.\end{array}\right.$$
\end{lemma}
Combining (\ref{eqw}) and (\ref{slopefull}) proves the Lemma after a few simple manipulations.

Altogether, in the case of fewer receive antennas than transmit antennas, MMSE equalization is not effective at high SNR. CS detection can compensate for the lack of receive antennas by setting its memory equal to the difference between the antenna numbers. The trade-off between complexity and performance is clear. This also explains why the rates saturate in Figure \ref{fig2}. In order for the asymptotic slope to be non-zero, the receiver memory must equal the number of eigenvalues of the channel matrix that equal zero. However, the asymptotic eigenvalue distribution of an ISI matrix is given by its frequency response $|H(\omega)|^2$.  The response (\ref{isic}) specifies a continuous band of zeros which means that any finite memory $K$ will ultimately yield an asymptotic slope that is also zero.

Let us now return to the special case of a block diagonal structure of $\Gr$. In this case we can show
\begin{lemma} \label{blocklm}
With a block diagonal structure of $\Gr$ the asymptotic slope of the ergodic rate becomes
$$S_{\infty}^{\mathrm{BD}}=\left\{\begin{array}{ll}\nt, & \nr\geq \nt \\ \sum_{m:K_m>\nt-\nr}K_m-(\nt-\nr),& \nt>\nr.\end{array}\right.$$
\end{lemma}
It can be verified that whenever $\max_m K_m\leq K$, the slope in Lemma \ref{cslemma} is always superior to the slope in Lemma \ref{blocklm}. 

With full complexity detection, one can show that \cite{MCT}
$$L_{\infty}^{\mathrm{full}} = \log(\nt)-\frac{\mathcal{J}(\nr,\nt)}{S_{\infty}},$$
where
$$\mathcal{J}(\nr,\nt)=\left\{\begin{array}{ll} \mathbb{E}\left[\log\left(\det\left( \vec{H}\vec{H}^{\mathrm{H}}\right)\right)\right], & \nr< \nt \\ \mathbb{E}\left[\log\left(\det\left( \vec{H}^{\mathrm{H}}\vec{H}\right)\right)\right], & \nr\geq\nt. \end{array}\right.$$
For  CS detection, we obtain 
\begin{lemma}
For a CS detector with memory $K$ whenever $S_{\infty}^{\mathrm{CS}}>0$, the high SNR power offset is
$$L_{\infty}^{\mathrm{CS}} = \log(\nt)-\frac{\mathcal{J}(\nr,\nt)}{S_{\infty}^{\mathrm{CS}}}
+(\nt-K)\frac{\mathcal{J}(\nr,\nt-K-1)}{S_{\infty}^{\mathrm{CS}}}-(\nt-K-1)\frac{\mathcal{J}(\nr,\nt-K)}{S_{\infty}^{\mathrm{CS}}}.$$
\end{lemma}
The Lemma is proved by inserting (\ref{eqw}) into (\ref{linf}) and taking the necessary limits.
If we compare $L_{\infty}^{\mathrm{full}}$ with $L_{\infty}^{\mathrm{CS}}$, we get that whenever $S_{\infty}^{\mathrm{CS}}>0$
$$L_{\infty}^{\mathrm{CS}}=L_{\infty}^{\mathrm{full}}+ \frac{1}{S_{\infty}^{\mathrm{CS}}}\sum_{\ell=0}^{\nt-K-1}\psi(\nr-\ell)-\frac{\nt-K}{S_{\infty}^{\mathrm{CS}}}\psi(\nr-\nt+K+1),$$
where $\psi(\cdot)$ is the digamma function
$$\psi(j)=\left\{\begin{array}{ll}\sum_{k=1}^{j-1} \frac{1}{k}-\gamma, &j\geq 1 \\ -\gamma, & j=1,\end{array} \right.$$
and $\gamma\approx 0.5772$ is the Euler-Mascheroni constant. This result follows directly from the observation \cite{pap39} 
$$\mathcal{J}(\nr,\nt)=\log_2(\exp(1))\sum_{\ell=0}^{\nt-1}\psi(\nr-\ell).$$

We turn to examples next,
\begin{example}
Let $\nt=6$ and $\nr=4$. In Figure \ref{f2} we plot the ergodic rates $\mathbb{E}[I_{\mathrm{GMI}}]$ against SNR for different values of the memory of the CS detector (the bottom curve in the legend of Figure \ref{f2} shall be discussed in Example 2).
As we can see, whenever the memory $K$ equals the difference $K=\nt-\nr=2$, the slope $S_{\infty}^{\mathrm{CS}}$ is optimal, i.e., $S_{\infty}^{\mathrm{CS}}=S_{\infty}^{\mathrm{full}}$. 
\end{example}

In our next example, we consider the block diagonal special case in Section \ref{bdcs}.
\begin{example}
For the same parameter setup as in Example 1, i.e., $\nt=6$ and $\nr=4$ let us consider a block diagonal structure of $\Gr$. The matrix $\Gr$ has dimensions $\nt\times\nt =6\times 6.$ What options do we have to select the block sizes? Clearly we can choose three blocks, i.e., $M=3$, and each block would then be $2\times 2$, i.e., $K_1=K_2=K_3=2$. However, we then always have $K_k<\nt-\nr=2$ so from Lemma 4 we get that $S_{\infty}^{\mathrm{BD}}=0$. The conclusion of this is that although the block-diagonal detector with $M=2$ is more complex than an MMSE equalizer, it does not improve much upon MMSE equalization at high SNR since $S_{\infty}^{\mathrm{BD}}=0$. 

Another choice would be to pick $M=2$ and use $K_1=K_2=3$. In view of Lemma 4, we now have $K_1=K_2>\nt-\nr$, and therefore we have that 
$$S_{\infty}^{\mathrm{BD}}=K_1-(\nt-\nr)+K_2-(\nt-\nr)=2.$$
This is still inferior to the slope of a CS detector with $K=2$.
An illustration of this rate is shown in Figure \ref{f2} and corresponds to the bottom curve in the legend of the figure.
\begin{figure}[h]
\begin{center}
\scalebox{1}{\includegraphics*{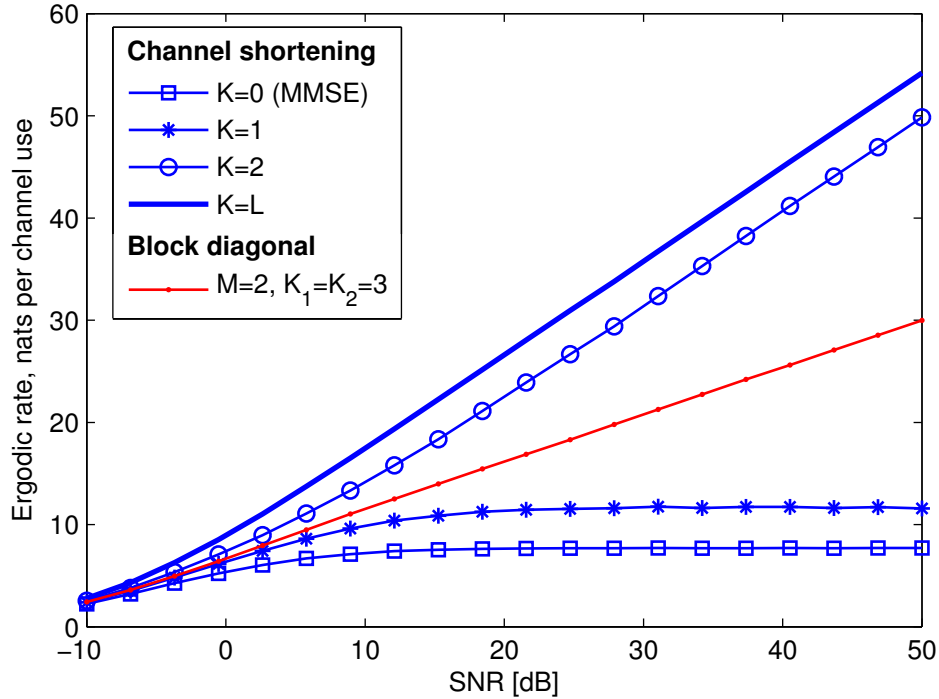}}
\vspace*{-3mm}
\caption{\label{f2} Ergodic rates of $4\times 6$ MIMO with IID complex Gaussian channel elements with CS detection of various memories  and with a block diagonal $\Gr$  with $K_1=K_2=3$. }
\end{center}
\vspace*{-3mm}  
\end{figure}
Note that in this case, the detection complexity of the block diagonal structure is lower than that of a CS with $K=2$. In the former case, we have two search \emph{trees} of depth 3, while in the latter case we have one \emph{trellis} of memory 2 with depth 6. However, performance is grossly reduced at high SNR.
\end{example}

In our next example we change the parameter settings.
\begin{example}
Let $\nt=6$ and $\nr=5$. In this case we already know from Lemma 3 that CS detection with $K=1$ is sufficient to reach $S_{\infty}^{\mathrm{CS}}=\min(\nr,\nt)=\nr=5$. 
Detection can be made on the basis of a trellis with memory $K=1$. The number of states  in the trellis is $|\mathcal{A}|$.

 For the block diagonal structure, we still have the two choices $M=3,\,K_1=K_2=K_3=2$ and $M=2,\,K_1=K_2=3.$ For $M=2$, we get from Lemma 4 that $S_{\infty}^{\mathrm{BD}}=3$, while for $M=3$ we get $S_{\infty}^{\mathrm{BD}}=4$. Hence, both cases are worse than CS with $K=1$. Complexity wise, the $M=2$ case is less complex than the CS $K=1$ case. However, for $M=3$ we have two trees with depth 3. The number of leaf nodes becomes $|\mathcal{A}|^3$ and this is one order worse than the number of states in the trellis multiplied with its branching number $|\mathcal{A}|$.
 An illustration of the discussed slopes is provided in Figure \ref{f3}.
\begin{figure}[h]
\begin{center}
\scalebox{1}{\includegraphics*{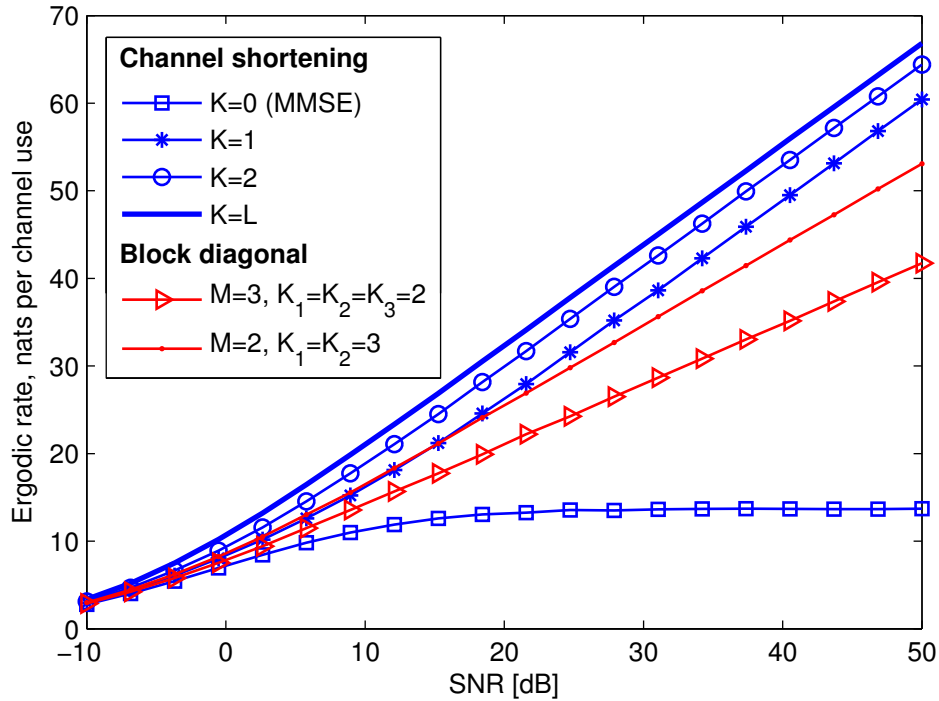}}
\vspace*{-3mm}
\caption{\label{f3} Illustration of the ergodic rates discussed in Example 3. }
\end{center}
\vspace*{-3mm}  
\end{figure}
\end{example}

We conclude by giving an illustration of what the optimized matrices $\Gr$ may actually look like.
\begin{example} 
Assume a channel matrix equal to 
$$\vec{H}=\left[\begin{array}{cccc} 0& 1&1&1 \\1&1&2&1\\1&-1&0&0 \end{array}\right]$$
and that $N_0=1$.
An optimized CS receiver with $K=1$ has
$$\Gr=\left[\begin{array}{cccc} 1.33& -1&0&0 \\-1&1.93&1.25&0\\0&1.25&2.04&0.83\\0&0&0.83&0.67 \end{array}\right].$$
The trellis structure is arising since there is no cross-coupling between the symbols $(x_1,x_3)$, $(x_1,x_4)$, and $(x_2,x_4)$. 
Now consider the block diagonal structure with $M=2$. After optimization, for example via the proof of Lemma 1, we get
$$\Gr=\left[\begin{array}{cccc} 1.33& -1&0&0 \\-1&1.33&0&0\\0&0&1.417&0.83\\0&0&0.83&0.67 \end{array}\right].$$
It is interesting to observe that the first and the last rows are not altered compared with the CS $K=1$ case. This is so since the memory is still 1 at these two rows even with the block diagonal structure. At the two middle rows, the cross-coupling between symbols $(x_2,x_3)$ has been broken and this enforces a somewhat "weaker" matrix $\Gr$ at these two rows. Further, due to the separated blocks, the trellis collapses into two trees.
Finally, note that the matrix $\Gr$ for the CS case is indefinite which means that the framework used in \cite{FM73} will not be able to produce this particular receiver setting. The block diagonal $\Gr$ is always positive semi-definite due to Property 2.
\end{example}

\section{Conclusion}
In this paper we have investigated rate optimized channel shortening receivers. We have shown that an optimized receiver can reach the chain rule of mutual information, up to the reduced memory assumed by the receiver. Further, we have shown that the formula for the achievable rate of a receiver with memory $K$ is essentially the same as for an MMSE equlizer of a MIMO system with $K$ transmit antennas less. This results enables significant analytical treatment. As an example of this, we derived the  capacity slope and the power offset at high SNR, and we demonstrated that receiver memory can compensate for a lack of receive antennas.

We also discussed that the classical model for channel shortening is bounded away from the optimal solution due to an inappropriate system model. A better model should be based upon Ungerboeck's formulation of trellis detection for ISI channels. The rate penalty for the classical model is, however, small, but closed form solutions for the optimal receiver parameters are only available for the Ungerboeck based CS framework.

\section*{Appendix A: Proof of Theorem \ref{thm1}} \label{proofthm1} 
From \cite{RP12} we have that after inserting the optimal $\Hr^{\mathrm{opt}}$, that the function to be optimized is 
\be \label{thm1eq1}I_{\mathrm{GMI}}(\Gr) = \log\left(\det\left(\vec{I}_{\nt}+\Gr\right)\right)-\mathrm{Tr}\left(\left(\vec{I}_{\nt}+\Gr\right)\vec{B}\right)+\nt.\ee
In \cite{RP12} this function was optimized by using the factorization $\vec{L}\vec{L}^{\mathrm{H}}=\left(\vec{I}_{\nt}+\Gr\right)$ where $\vec{L}$ is a lower triangular matrix with only the first $K+1$ off diagonals holding non-zero elements. The first step of the proof of Theorem \ref{thm1} is to redo the derivations from \cite{RP12}, but with the alternative factorization $\vec{U}\vec{D}\vec{U}^{\mathrm{H}}=\left(\vec{I}_{\nt}+\Gr\right)$, where $\vec{U}$ is an upper triangular matrix with ones along the main diagonal and $\vec{D}$ is a diagonal matrix. The derivations to find the optimal $\vec{U}$ and $\vec{D}$ are identical to those for finding the optimal $\vec{L}$ in \cite{RP12}, and we therefore only state the final result. 
Since $\vec{U}$ is upper triangular with only the first $K+1$ off diagonals not equal to 0, it has the structure
$$\tilde{\vec{U}}=\left[\begin{array}{cccc} \boxed{\begin{array}{ccc} && \\ & \vec{U}_1 &\\ && \end{array}}& \boxed{\begin{array}{c} \;\\ \!\!\!\!\vec{u}_{K\!+\!1}\!\!\!\! \\ \; \end{array}}&  & \begin{array}{c} \\ \vec{0} \\ \end{array}		\\ & \begin{array}{c} 1 \\ \\  \end{array}& \begin{array}{c} \ddots \\ \\ \ddots \end{array} & \boxed{\begin{array}{c} \;\\ \!\!\!\!\vec{u}_{\nt}\!\!\!\! \\ \; \end{array}}	\\ \vec{0} & & & 1	\end{array}\right].$$
The optimal $\vec{U}$ has
$$\vec{u}_k=-\left(\vec{B}_{k-1}^{k-K}\right)^{-1}\vec{b}_{(k-K):(k-1),k},$$
where $\vec{b}_{i:j,\ell} =\left[B_{i,\ell} \; B_{i+1,\ell} \;\ldots\; B_{j,\ell}\right]^{\mathrm{T}}.$
Further, $\vec{U}_1$ is the "`U"'-matrix in  an UDL factorization of $(\vec{B}_{K}^1)^{-1}$. 
The optimal matrix $\vec{D}$ has the structure
$$\vec{D}=\mathrm{diag}\left[d_1,d_2,\ldots,d_{\nt}\right],$$
where $$d_k=\frac{\det\left( \vec{B}_{k-1}^{k-K}\right)}{\det\left( \vec{B}_{k}^{k-K}   \right)}, \quad k>K,$$ and where $d_1,d_2,\ldots,d_K$ are the diagonal elements of the "`D"'-matrix in an UDL factorization of $(\vec{B}_{K}^1)^{-1}$. 

The formulas for the optimal $\vec{U}$ and $\vec{D}$ matrices coincide with \cite[Eqs. (18)-(22)]{KM00}. From \cite[Proof of Lemma 1]{KM00} (see also \cite[Section II-C]{KM00}), this implies immediately that $(\vec{I}_{\nt}+\Gr^{\mathrm{opt}})^{-1}$ is the $K$-band extension of $\vec{B}$ which concludes the proof.

\section*{Appendix B: Proof of Lemma1} \label{proofLemma1} 
Since we know from Theorem \ref{thm1} that $(\vec{I}_{\nt}+\Gr^{\mathrm{opt}})^{-1}$ is the $K$-band extension of $\vec{B}$, the corollary is a trivial consequence of \cite[Theorem 2]{KM00}. 
An alternative proof appears in \cite{Medhat}.
\section*{Appendix C: A sharpended version of \cite[Theorem 2]{KM00}}
Let us first state \cite[Theorem 2]{KM00},
\begin{theorem}[Theorem 2 in \cite{KM00}]For an arbitrary square matrix $\vec{C}$, there exist a unique matrix $\vec{R}$ such that $\mathrm{diag}_K(\vec{C})=\mathrm{diag}_K(\vec{R})$, and a matrix $\vec{A}$ that equals zero along the center $2K+1$ diagonals, i.e., $\vec{A}$ satisfies $\vec{A}=\vec{A}-\mathrm{diag}_K(\vec{A})$ such that
$$\vec{C}=\vec{R}+\vec{A}\;\mathrm{and}\;\vec{R}^{-1}=\mathrm{diag}_K(\vec{R}^{-1}).$$
The matrix $\vec{R}$ is the $K$-band extension of $\vec{C}$. 
\end{theorem}

We can sharpen this theorem as follows
\begin{lemma} \label{applemme}
If the matrix $\vec{C}$ in \cite[Theorem 2]{KM00} is positive definite, so is the matrix $\vec{R}^{-1}$.
\end{lemma}
The proof of Lemma \ref{applemme} is a simple identification of terms. We can identify the matrix $\vec{C}$ in \cite[Theorem 2]{KM00} by the matrix $\vec{B}$ which is always positive definite. The matrix $\vec{R}^{-1}$ is identified by $\vec{I}_{\nt}+\Gr^{\mathrm{opt}}$ and this is a positive definite matrix by construction. This concludes the proof.

\section*{Appendix D: Proof of Corollary 2} \label{proofLemma2}
From (\ref{thm1eq1}), we have that the statement of the corollary is equivalent to proving that
$$\mathrm{Tr}\left(\left(\vec{I}_{\nt}+\Gr^{\mathrm{opt}}\right)\vec{B}\right)=\nt.$$
From Corollary \ref{inversechar}, we know that we can express $\vec{B}$ as
$$\vec{B}=(\vec{I}_{\nt}+\Gr^{\mathrm{opt}})^{-1}+\vec{A}$$
where $\vec{A}=\vec{A}-\mathrm{diag}_K(\vec{A})$. Hence, 
$$(\vec{I}_{\nt}+\Gr^{\mathrm{opt}})\vec{B}=\vec{I}_{\nt}+(\vec{I}_{\nt}+\Gr^{\mathrm{opt}})\vec{A}.$$
Since $\Gr^{\mathrm{opt}}+\vec{I}_{\nt}=\mathrm{diag}_K(\vec{I}_{\nt}+\Gr^{\mathrm{opt}})$ we have that 
$$\mathrm{Tr}\left((\vec{I}_{\nt}+\Gr^{\mathrm{opt}})\vec{B}\right)=\mathrm{Tr}\left(\vec{I}_{\nt}+(\vec{I}_{\nt}+\Gr^{\mathrm{opt}})\vec{A}\right)=\mathrm{Tr}\left(\vec{I}_{\nt}\right)=\nt.$$
This concludes the proof.

\section*{Appendix E: Proof of Theorem 2} \label{proofthm2}
From Corollary \ref{corsimple} we have that 
$$I_{\mathrm{GMI}}(\Hr^{\mathrm{opt}},\Gr^{\mathrm{opt}})=\log\left(\det\left(\vec{I}_{\nt}+\Gr^{\mathrm{opt}}\right)\right).$$ 
The factorization $\vec{I}_{\nt}+\Gr^{\mathrm{opt}}=\vec{U}\vec{D}\vec{U}^{\mathrm{H}}$ in the proof of Theorem \ref{thm1}, yields
$$I_{\mathrm{GMI}}(\Hr^{\mathrm{opt}},\Gr^{\mathrm{opt}})=\log\left(\det\left(\vec{D}\right)\right).$$
Let us now focus on $\det(\vec{D})$. From the proof of Theorem \ref{thm1}, this equals,
\bea \label{kanske} 
\det(\vec{D}) &=& \prod_{k=1}^{\nt}d_k \nonumber \\
&=&\prod_{k=1}^{K}d_k \prod_{k=K+1}^{\nt}\frac{\det\left(\vec{B}_{k-1}^{k-K}\right)}{\det\left(\vec{B}_{k}^{k-K}\right)} \nonumber \\
&=& \frac{1}{\det\left(\vec{B}_{K}^{1}\right)}\prod_{k=K+1}^{\nt}\frac{\det\left(\vec{B}_{k-1}^{k-K}\right)}{\det\left(\vec{B}_{k}^{k-K}\right)}.
\eea 
Let $\bar{\vec{H}}_{[k,n]}$ denote the same matrix as $\vec{H}$, but with columns $k,k+1,\ldots,n$ removed, and $\vec{H}_{[k,n]}$ denote the matrix $\vec{H}$ with all columns except $k,k+1,\ldots,n$ removed. Then, each term $\det\left(\vec{B}_{n}^{k}\right)$ equals
\bea \label{kanske2}
\det\left(\vec{B}_{n}^{k}\right) &=& \det\left(\vec{I}_{\nt}-\frac{\vec{H}_{[k,n]}^{\mathrm{H}}}{\sqrt{N_0}}\left(\vec{I}_{\nt}+\frac{\vec{H}\vec{H}^{\mathrm{H}}}{N_0}\right)^{-1}  \frac{\vec{H}_{[k,n]}}{\sqrt{N_0}}\right) \nonumber \\
&=&\det\left(\vec{I}_{\nt}-\frac{\vec{H}_{[k,n]}\vec{H}_{[k,n]}^{\mathrm{H}}}{N_0}\left(\vec{I}_{\nt}+\frac{\vec{H}\vec{H}^{\mathrm{H}}}{N_0}\right)^{-1} \right) \nonumber \\
&=&\det\left(\left(\vec{I}_{\nt}-\frac{\vec{H}_{[k,n]}\vec{H}_{[k,n]}^{\mathrm{H}}}{N_0}+\frac{\vec{H}\vec{H}^{\mathrm{H}}}{N_0}\right)\left(\vec{I}_{\nt}+\frac{\vec{H}\vec{H}^{\mathrm{H}}}{N_0}\right)^{-1} \right) \nonumber \\
&=&\det\left(\left(\vec{I}_{\nt}+\frac{\bar{\vec{H}}_{[k,n]}\bar{\vec{H}}_{[k,n]}^{\mathrm{H}}}{N_0}\right)\left(\vec{I}_{\nt}+\frac{\vec{H}\vec{H}^{\mathrm{H}}}{N_0}\right)^{-1} \right) \nonumber \\
&=& \frac{\det\left(\vec{I}_{\nt}+\frac{\bar{\vec{H}}_{[k,n]}\bar{\vec{H}}_{[k,n]}^{\mathrm{H}}}{N_0}\right)}{\det\left(\vec{I}_{\nt}+\frac{{\vec{H}}{\vec{H}}^{\mathrm{H}}}{N_0}\right)}.
\eea
Inserting  (\ref{kanske2}) into (\ref{kanske}) yields
\bea \label{kanske3}
\log(\det(\vec{D}))&=&\log\left(\det\left(\vec{I}_{\nt}+\frac{\vec{H}\vec{H}^{\mathrm{H}}}{N_0}\right)\right)-\log\left(\det\left(\vec{I}_{\nt}+\frac{\bar{\vec{H}}_{[1,K]}\bar{\vec{H}}_{[1,K]}^{\mathrm{H}}}{N_0}\right)\right)\nonumber \\ &&+\sum_{k=K+1}^{\nt}\log\left(\det\left(\vec{I}_{\nt}+\frac{\bar{\vec{H}}_{[k-K,k-1]}\bar{\vec{H}}_{[k-K,k-1]}^{\mathrm{H}}}{N_0}\right)\right)\nonumber \\
&&-\sum_{k=K+1}^{\nt}\log\left(\det\left(\vec{I}_{\nt}+\frac{\bar{\vec{H}}_{[k-K,k]}\bar{\vec{H}}_{[k-K,k]}^{\mathrm{H}}}{N_0}\right)\right).
\eea
By introducing the notation $\vec{G}_{[k,n]}=\bar{\vec{H}}_{[k,n]}\bar{\vec{H}}_{[k,n]}^{\mathrm{H}}$, a change of variable in the summations ($k\to k-K$), and canceling the second term of the right-hand-side of (\ref{kanske3}) with the first term of the first sum, gives the statement of the theorem.

\section*{Appendix F: Proof of Corollary \ref{chain}} \label{proofchain}
With the system model (\ref{sysmod}), we have that
$$\log\left(\det\left(\vec{I}_{\nt}+\frac{\vec{H}\vec{H}^{\mathrm{H}}}{N_0}\right)\right)=h(\vec{Y})-h(\vec{N})$$
where $h(\cdot)$ is the differential entropy operator.
Similarly, 
$$\log\left(\det\left(\vec{I}_{\nt}+\frac{\vec{G}_{[k,n]}}{N_0}\right)\right)=h(\vec{Y}|X_n,X_{n-1},\ldots,X_k)-h(\vec{N}).$$
Inserting these two identities into Theorem \ref{main} gives the statement of the corollary after a few manipulations.

\section*{Appendix G: Proof of Lemma \ref{block}} \label{blockproof}
The proof is straightforward. Let us consider a specific block, say the first.
The input-output relation is
$$\vec{y}=\vec{H}_{[1,K_1]}\vec{x}_1+\vec{w}_1$$
where $\vec{x}_1=[x_1,x_2,\ldots,x_{K_1}]^{\mathrm{T}}$ and $\vec{w}_1\sim\mathcal{CN}\left(\vec{0},N_0\vec{I}_{\nr}+\bar{\vec{H}}_{[1,K_1]}\bar{\vec{H}}_{[1,K_1]}^{\mathrm{H}}\right)$. The first block of $\Gr$ has dimension $K_1\times K_1$, which means that full complexity detection of $\vec{x}_1$ is performed. Therefore, the achievable rate for the first block becomes precisely
$$I(\vec{Y};\vec{X}_1)=\sum_{k=1}^{K_1}I(\vec{Y};X_k|X_{k-1},\ldots,X_1).$$
Similar arguments hold for the remaining blocks, and summing the rates over all blocks gives the statement of the Lemma.

\end{document}